\documentclass[conference]{IEEEtran}
\IEEEoverridecommandlockouts
\usepackage{cite}
\usepackage{amsmath,amssymb,amsfonts}
\usepackage{algorithmic}
\usepackage{graphicx}
\usepackage{textcomp}
\usepackage{xcolor}
\usepackage{hyperref}
\usepackage{bm}

\def\BibTeX{{\rm B\kern-.05em{\sc i\kern-.025em b}\kern-.08em
    T\kern-.1667em\lower.7ex\hbox{E}\kern-.125emX}}
\begin{document}

\title{Faster-than-Nyquist Equalization with Convolutional Neural Networks\\
}

\author{\IEEEauthorblockN{Bruno De Filippo\IEEEauthorrefmark{1}, Carla Amatetti\IEEEauthorrefmark{1}, Alessandro Vanelli-Coralli\IEEEauthorrefmark{1}}
\IEEEauthorblockA{\IEEEauthorrefmark{1}Department of Electrical, Electronic, and Information Engineering (DEI), Univ. of Bologna, Bologna, Italy}
\{bruno.defilippo, carla.amatetti2, alessandro.vanelli\}@unibo.it}

\maketitle

\begin{abstract}
Faster-than-Nyquist (FTN) signaling aims at improving the spectral efficiency of wireless communication systems by exceeding the boundaries set by the Nyquist-Shannon sampling theorem. 50 years after its first introduction in the scientific literature, wireless communications have significantly changed, but spectral efficiency remains one of the key challenges. To adopt FTN signaling, intersymbol interference (ISI) patterns need to be equalized at the receiver. Motivated by the pattern recognition capabilities of convolutional neural networks with skip connections, we propose such deep learning architecture for ISI equalization and symbol demodulation in FTN receivers. We investigate the performance of the proposed model considering quadrature phase shift keying modulation and low density parity check coding, and compare it to a set of benchmarks, including frequency-domain equalization, a quadratic-programming-based receiver, and an equalization scheme based on a deep neural network. We show that our receiver outperforms any benchmark, achieving error rates comparable to those in additive white Gaussian noise channel, and higher effective throughput, thanks to the increased spectral efficiency of FTN signaling. With a compression factor of 60\% and code rate 3/4, the proposed model achieves a peak effective throughput of 2.5 Mbps at just 10dB of energy per bit over noise power spectral density ratio, with other receivers being limited by error floors due to the strong intersymbol interference. To promote reproducibility in deep learning for wireless communications, our code is open source at the repository provided in the references.
\end{abstract}

\begin{IEEEkeywords}
Faster-Than-Nyquist Signaling, Inter-Symbol Interference, Deep Learning, Convolutional Neural Networks
\end{IEEEkeywords}

\section{Introduction}\label{ch:1_intro}
\noindent Since the dawn of wireless communications, one of the greatest challenges has always been how to improve the radio resource utilization. As spectrum is a scarce and costly asset, both academia and industrial researchers have proposed new techniques over the years to increase the number of bits per second that can be transmitted over a fixed bandwidth, \textit{e.g.}, cell-free multiple input multiple output (CF-MIMO) \cite{bib:CFMIMO} and non-orthogonal multiple access (NOMA) \cite{bib:NOMA}. The appeal of faster-than-Nyquist (FTN) signaling lies in the straightforwardness of its approach: to achieve a higher data rate, one can simply transmit at a faster symbol rate using the same pulse. In 1975, J. E. Mazo investigated the effect of intersymbol interference (ISI) on symbols transmitted at a rate $R_s$ higher than that allowed by the famous Nyquist-Shannon sampling theorem, \textit{i.e.}, twice the bandwidth of the modulating pulse. The author observed that, for ideal sinc pulses with bandwidth $W$ and binary transmissions, no significant reduction in minimum Euclidean distance (MED) between received symbol sequences could be observed as long as $\frac{2W}{R_s}\geq \tau_{Mazo}=0.802$, the so-called "Mazo limit" \cite{bib:mazo}. This result has motivated several authors to investigate FTN and push the technique over the Mazo limit, \textit{e.g.}, \cite{bib:andersonNovel,bib:andersonRx,bib:reducedComp,bib:FDE,bib:SDR}. The scientific community has focused on identifying and exploiting the intricate ISI patterns for equalizing the received FTN data streams. Today, 50 years after Mazo's original paper, deep learning (DL) techniques are pushing the boundaries of FTN transmissions. In \cite{bib:DNNwindow}, a deep neural network (DNN) was trained to equalize FTN quadrature amplitude modulation (QAM) symbols with a sliding window mechanism. The input of the DNN spans more symbols than those being equalized, leading to a more accurate equalization and, thus, lower block error rate (BLER), at the expense of a slight increase in complexity. To identify the temporal patterns introduced by FTN, recurrent neural networks (RNNs) in FTN were investigated in \cite{bib:LSTM} and \cite{bib:RNN}. On the one hand, \cite{bib:LSTM} implemented both uni- and bi-directional long-short term memory (LSTM) models to demodulate binary phase shift keying (BPSK) symbols, observing a reduction in the Bit Error Rate (BER) floor when using the second model over the first. On the other hand, the authors in \cite{bib:RNN} used a more traditional RNN to equalize FTN BPSK symbols under strong ISI, \textit{i.e.}, with a compression factor $\tau=0.5$.
\noindent The DL literature has grown exponentially in the past decade, with image processing and pattern recognition being among the fundamental areas of research. Inspired by the great performance of convolutional neural networks (CNNs) with skip connections in image processing, \textit{e.g.}, \cite{bib:pattern}, we investigate whether such results apply to the physical layer of wireless communication systems, too. CNNs have already been successfully employed in studies related to physical layer techniques, \textit{e.g.}, for NOMA \cite{bib:NOMA} and channel prediction \cite{bib:channelPrediction}. To the best of our knowledge, their application to FTN has only been proposed to assist the sum-product detection algorithm in \cite{bib:SPA}, where CNNs act as function nodes for the FTN factor graph, approaching the BER analytical bound for convolutionally-coded and BPSK-modulated transmissions. A comprehensive approach to FTN equalization with CNNs, exploiting their pattern recognition capabilities to the fullest, has not been evaluated yet. Motivated by these considerations and analyses, in this paper we:
\begin{itemize}
    \item Propose the application of CNNs with skip connections in FTN receivers to identify and equalize ISI patterns;
    \item Evaluate and compare the BER, BLER, and throughput (TP) achieved by the proposed model against several benchmarks, including a DL-based equalization scheme.
\end{itemize}
In addition, in an effort to advocate for reproducibility, taking once again inspiration from the DL scientific community, we open-source our code in a public repository \cite{bib:repo}.

\section{System model}\label{ch:2_systemModel}
\begin{figure}[t]
    \centerline{\includegraphics[width=7.5cm]{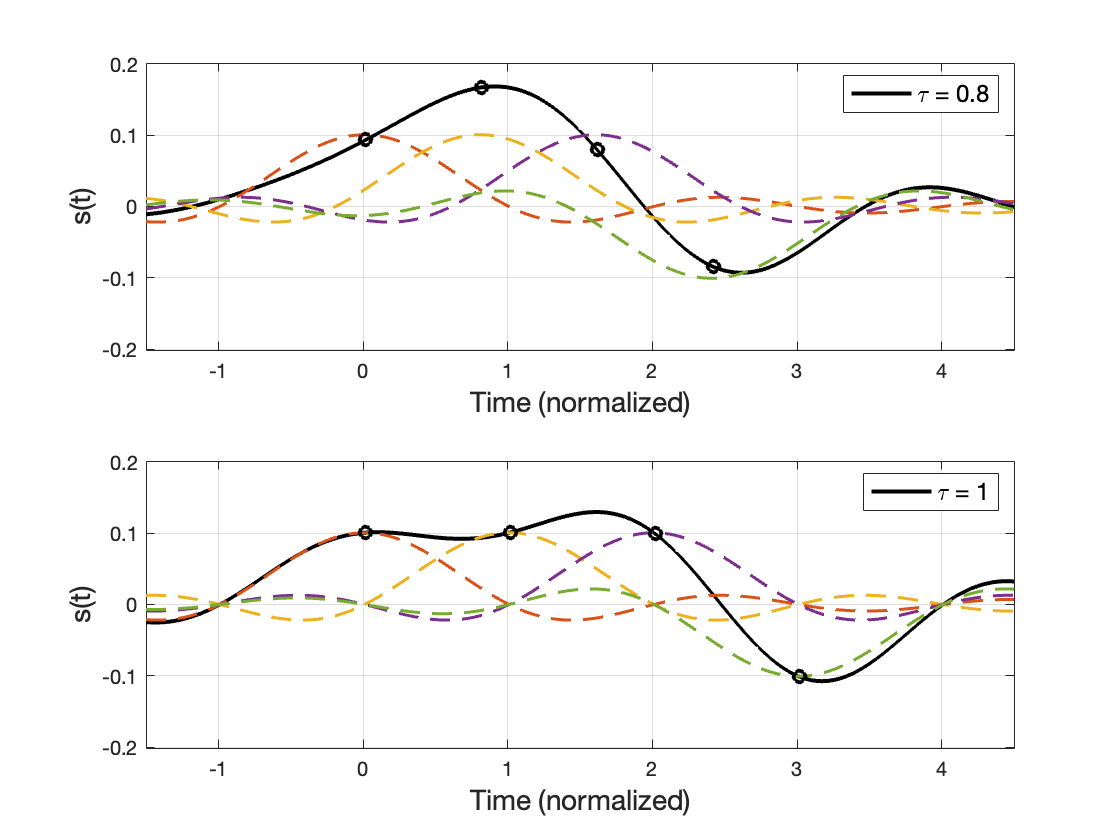}}
    \caption{Example of FTN baseband signal (above) vs at-Nyquist baseband signal (below) corresponding to 4 binary symbols and sinc shaping pulse.}
    \label{fig:signal}
\end{figure}

\noindent We here consider a typical transceiver, consisting of a bit source, a channel encoder, an interleaver, a digital modulator, and pulse shaping on the transmitter side, as well as a matched filter, a sampler at FTN rate $R_s$, a demodulator, a deinterleaver, and a channel decoder on the receiver side. Given the exploratory nature of this work, we consider additive white Gaussian noise (AWGN) as the only channel impairment. The focus of this paper is the demodulator, which is here implemented with a CNN.
The FTN baseband signal is modeled as follows. A sequence of energy-normalized quadrature phase shift keying (QPSK) symbols $\{x_n\}_{n\in \mathbb{Z}}$ is generated at the output of the digital modulator from low density parity check (LDPC)-coded bits and fed to the FTN pulse shaper with symbol time $T_s = \frac{1}{R_s} = \tau \frac{1}{2W} = \tau T_N$, where $0 < \tau < 1$ is the FTN compression factor and $T_N$ is the symbol time at Nyquist rate. The filter employs the unit-energy square root raised cosine (SRRC) pulse $h(t)$ with roll-off factor $\beta=0.5$ and bandwidth $W$, producing the following signal \cite{bib:DNNwindow}:
\begin{equation}\label{eqn:txSignal}
    s(t)=\sum _{n=-\infty }^{+\infty }x_n h(t-n\tau T_N).
\end{equation}
Clearly, the strict upper constraint on $\tau$ results in a violation of the Nyquist-Shannon sampling theorem; thus, as previously anticipated, ISI is introduced (Figure \ref{fig:signal}). After being corrupted by AWGN, represented by the zero-mean circularly symmetric Gaussian random process $n(t)$ with variance $\frac{N_0}{2}$, with $N_0$ being the noise power spectral density, the signal is processed by a filter matched to $h(t)$. Let $g(t) = \int h(\tau)h^*(\tau-t)d\tau$ denote the ISI channel autocorrelation function, and $\eta(t) = \int n(\tau)h^*(\tau-t)d\tau$ the matched filter response to AWGN \cite{bib:FDE}, the output of the matched filter is:
\begin{equation}\label{eqn:rxSignal}
    y(t)= \sum _{n=-\infty }^{+\infty }x_{n}g(t-n\tau T_N)+\eta(t).
\end{equation}
Sampling at FTN rate $R_s$ yields a sequence $\{y_k\}_{k\in \mathbb{Z}}$, where each received symbol can be computed as
\begin{equation}\label{eqn:sampled}
    y_k= \sum _{n=-\infty }^{+\infty }x_{n}g((k-n)\tau T_N)+\eta(k\tau T_N),
\end{equation}
or, considering a block of $N_s$ symbols $\mathbf{y}=\{ y_k\}_{1, \ldots, N_s}$,
\begin{equation}\label{eqn:matricial}
    \mathbf{y}= \mathbf{Gx} + \bm{\eta}.
\end{equation}
In \eqref{eqn:matricial}, $\mathbf{G}$ represents the ISI channel matrix, \textit{i.e.}, the contribution of preceding and succeeding symbols on the current one due to ISI. We truncate the channel autocorrelation function samples $\{g_n\}_{n\in\mathbb{Z}}=g(n\tau T_N)$ to the centermost $2L_{I}+1 < N_s$ taps, where $L_{I}$ represents the one-sided ISI span, resulting in $\mathbf{G}$ being structured as a $N_s \times N_s$ Toeplitz matrix with $2\times(N_s - L_{I} - 1)$ zero diagonals \cite{bib:reducedComp}:
\begin{equation}
    \mathbf{G}=\begin{bmatrix}
                    1         & g_{-1}    & \cdots & g_{-L_{I}} & 0          & 0      & \cdots \\
                    g_{1}     & 1         & \ddots & \ddots     & g_{-L_{I}} & 0      & \ddots \\
                    \vdots    & \ddots    & \ddots & g_{-1}     & \ddots     & \ddots & \ddots \\
                    g_{L_{I}} & \ddots    & g_{1}  & 1          & g_{-1}     & \ddots & \ddots \\
                    0         & g_{L_{I}} & \ddots & g_{1}      & 1          & \ddots & \ddots \\
                    0         & 0         & \ddots & \ddots     & \ddots     & \ddots & \ddots \\
                    \vdots    & \ddots    & \ddots & \ddots     & \ddots     & \ddots & \ddots \\
                \end{bmatrix}
\end{equation}
Note that the structure of $\mathbf{G}$ does not fully capture ISI at the boundaries, \textit{e.g.}, $y_1$ is affected only by the upcoming symbols in the block and not by previous transmissions; hence, this effect should be taken into account when simulating a continuous stream of data. The random vector $\bm{\eta}\sim \mathcal{N}(\mathbf{0}, \frac{N_0}{2} \mathbf{G})$ represents the effect of AWGN after the matched filter, resulting in colored noise. While some studies have applied methods to whiten the noise, \textit{e.g.}, \cite{bib:reducedComp}, we chose not to further manipulate our samples. Thus, $\mathbf{y}$ is the input data of the proposed CNN.

\section{Proposed CNN-based FTN demodulator}\label{sec:3_CNN}
\begin{figure}[t]
    \centerline{\includegraphics[width=8cm]{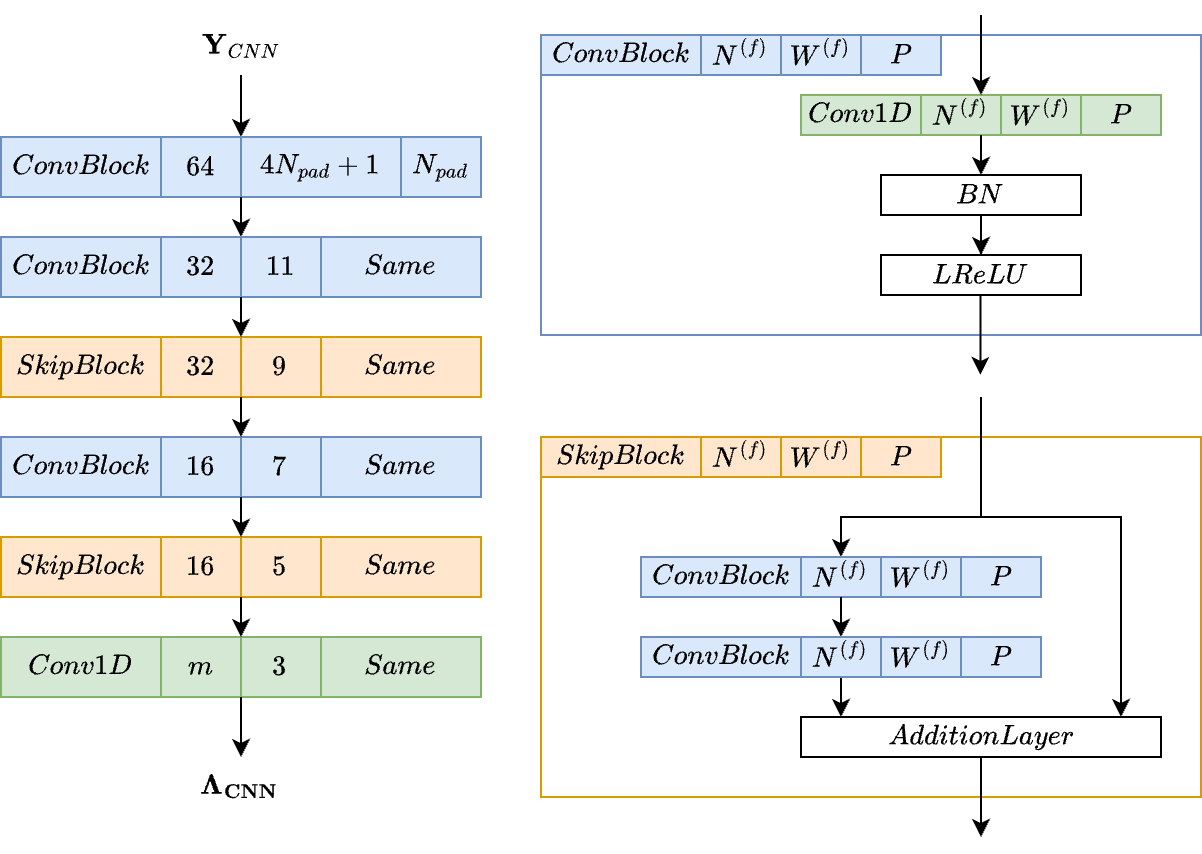}}
    \caption{Diagram of the proposed CNN.}
    \label{fig:CNNscheme}
\end{figure}

\noindent The task of the DL model is to convert a sequence of received symbols into the corresponding sequence of coded bit log-likelihood ratios (LLRs); thus, while the CNN must be tailored to a modulation order, it remains independent of the coding scheme. The receiver proposed in this work takes inspiration from pattern recognition models and employs a sequence of 1-dimensional convolutional (Conv1D) layers, leaky rectified linear unit (LReLU) activation functions, and batch normalization (BN) layers, together with skip connections; its structure is reported in Figure \ref{fig:CNNscheme}.
Each Conv1D layer includes $N^{(f)}$ filters with $W^{(f)}$ taps, where each filter is applied to the $N^{(f-1)}$ channels of the input tensor to generate a new sequence of length $L$; thus, the number of channels of the output tensor coincides with $N^{(f)}$. Its length depends on the amount $P$ of padding applied before filtering ($P=$"same" leaves the sequence length unchanged). LReLU activation functions introduce non-linearities in the CNN, allowing both positive and negative activations, differently from their non-leaky counterpart. BN layers normalize Conv1D activations, resulting in a more stable training process. Finally, skip connections allow Conv1D layers to extract features and apply them to their input tensor, rather than directly outputting the modified input; this enables the effective training of deeper CNNs.

\subsection{CNN input}\label{ch:3.1_CNNin}
\begin{figure}[t]
    \centerline{\includegraphics[width=6.5cm]{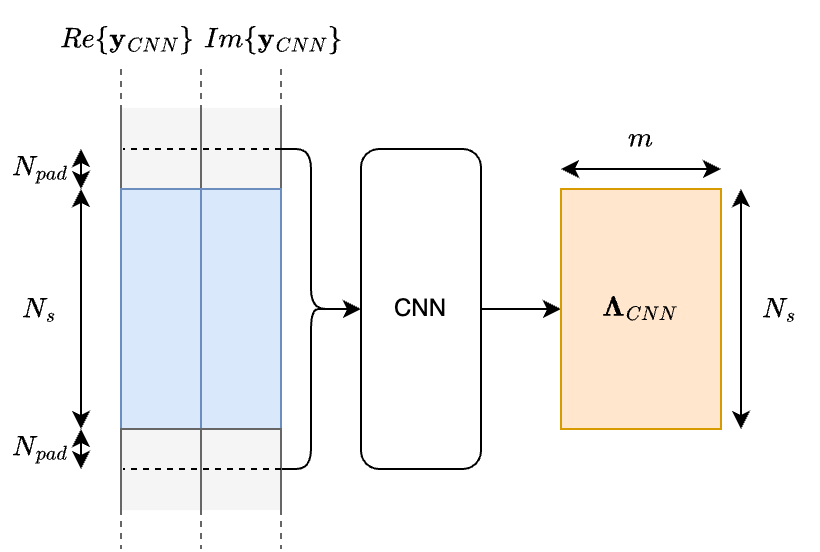}}
    \caption{CNN input pre-processing.}
    \label{fig:preprocess}
\end{figure}

\noindent The CNN pre-processing steps are depicted in Figure \ref{fig:preprocess}.
A block of $N_s$ received symbols $\mathbf{y}$ is padded on the left with the final $N_{pad}$ symbols of the preceding block and on the right with the initial $N_{pad}$ symbols of the succeeding block; we refer to such complex vector of length $L_{in} = N_s + 2N_{pad}$ as $\mathbf{y}_{CNN}$. While this method delays by $N_{pad}T_s$ the demodulation, it enhances the CNN's ability to extract ISI patterns. The vector is then split into its real and imaginary components, which are stacked together to form a $L_{in} \times 2$ matrix $\mathbf{Y}_{CNN}$.

\subsection{CNN output}\label{ch:3.2_CNNout}
\noindent To fully exploit the capabilities of the LDPC decoder, the CNN outputs the bit LLRs related to the data block of interest. We call such $N_s \times m$ matrix $\bm{\Lambda}_{CNN}$, where the $i$-th row contains the vector of $m$ LLRs associated with the $i$-th data symbol in the block. The corresponding $N_s \times m$ matrix of transmitted bits, used as labels during the training process, is named $\mathbf{B}$.

\subsection{Loss function}\label{ch:3.3_lossFunction}
\noindent While LLRs assume values in $\mathbb{R}$, it is still possible to approach the problem as a classification task. Conventionally, bit probabilities can be obtained at the output of a DL model using the binary crossentropy (BCE) loss function if the sigmoid activation $\sigma(x)=(1+e^{-x})^{-1}$ is applied to the last layer. Rather than post-processing the sigmoid output, the model can be trained to produce LLRs by incorporating the sigmoid activation into the loss function. This results in the so-called BCE with logits (L-BCE) loss function \cite{bib:NOMA}, which is computed by averaging the L-BCE between each LLR $\lambda_i \in \bm{\Lambda}_{CNN}$ and the corresponding true bit $b_i \in \mathbf{B}$, $L-BCE_i$:
\begin{equation}
    L-BCE_i =
        \begin{cases}
            ln\left(1 + e^{\lambda_i}\right) & \text{ if }b_i = 0 \\
            ln\left(1 + e^{-\lambda_i}\right) & \text{ if }b_i = 1
        \end{cases}
\end{equation}

\subsection{Dataset, training, and inference}\label{ch:3.4_DataTrainInfer}
\noindent The CNN is pretrained on a synthetic dataset with no predefined size, \textit{i.e.}, a batch of $N_B$ input matrices and their corresponding labels is generated during each epoch according to the system model presented in Section \ref{ch:2_systemModel}. The $E_s/N_0$ ratio, where $E_s =\mathbb{E}\left[\lvert x_n\rvert^2\right]$ is the average energy per symbol, is randomly set for each example within a batch. The loss function gradients with respect to the CNN parameters are computed with backpropagation; then, the Adam optimizer updates such parameters using L2 regularization. The learning rate is reduced at loss function plateaus and training is terminated via early stopping after long plateaus. We separately train one CNN per value of $\tau$. Once training is completed, the models can then be deployed for inference within the receiving chain; online training is beyond the scope of this work.

\section{Results}\label{ch:4_results}
\noindent In this section, we analyze the link level performance of the proposed receiver. Simulations were conducted in the MATLAB computing environment with the aid of the CVX package \cite{bib:cvx}; the source code is available at \cite{bib:repo}. Due to space constraints, only the main simulation parameters are reported in Table \ref{tab:parameters}. To ensure that boundary conditions in the ISI channel matrix $\mathbf{G}$ do not affect the validity of our results, we add further padding to each generated block of symbols, simulating a continuous stream; thus, each symbol fed to the CNN is equally affected by ISI.
\setlength{\tabcolsep}{5pt}
\renewcommand{\arraystretch}{1.1}
\begin{table}
    \centering
    \caption{Simulation parameters}
    \label{tab:parameters}
    \begin{tabular}{|c|c|}
        \hline
        \textbf{Parameter} & \textbf{Value} \\
        \hline
        Maximum Monte Carlo Iterations & $N_{MC} = 10^5$ \\
        \hline
        Modulation & QPSK ($m=2$) \\
        \hline
        Code rate & $R_c = \{1/2, 3/4\}$ \\
        \hline
        FTN compression & $\tau = \{0.6, 0.7\}$ \\
        \hline
        One-sided ISI span & $L_I = \{33, 28\}$ symbols\\
        \hline
        Nyquist symbol time & $T_N = 1\mu s$ \\
        \hline
        Block size & $N_s = 50$ symbols\\
        \hline
        Block padding & $N_{pad} = 12$ symbols\\
        \hline
        Batch size & $N_B = 4096$\\
        \hline
        Initial learning rate & $0.01$\\
        \hline
        Learning rate decay factor & $10$\\
        \hline
        Learning rate decay patience & $50$ epochs\\
        \hline
        Early stopping patience & $150$ epochs\\
        \hline
        L2 regularization factor & $10^{-4}$\\
        \hline
    \end{tabular}
\end{table}

\subsection{Benchmark receivers}\label{ch:4.1_benchmarks}
\noindent
For a complete evaluation, we chose a set of benchmark receivers from the literature to implement alongside our receiver:
\begin{itemize}
    \item MED \cite{bib:MED}: the optimal symbol-by-symbol QAM demodulator without any FTN-dedicated pre-processing.
    \item Frequency-domain equalization (FDE) \cite{bib:FDE}: a cyclic prefix (CP) of length $2\nu_{FDE}$ is added to the symbols block, transforming $\mathbf{G}$ in a circulant matrix. The minimum mean squared error (MMSE) detection algorithm is then applied on the Fourier transform of the received symbols.
    \item Set theory semi-definite-relaxation-based sequence estimation (STSDRSE) \cite{bib:SDR}: the FTN sequence detection task is modeled as a non-convex optimization problem, of which a relaxed version with quadratic constraint is solved. Gaussian randomization ensures the original constraints are satisfied. To achieve accurate BER plots with reasonable processing time, we limited the number of Monte Carlo iterations with this receiver to $10^4$.
    \item DNN \cite{bib:DNNwindow}: a DNN with 4 hidden layers for FTN equalization that outputs an ideally ISI-free sequence of symbols to be fed to the MED receiver. In a similar fashion to our receiver, a sliding window mechanism extracts a sequence of symbols with padding. It is important to note that the original paper does not specify all parameters required to ensure full reproducibility. \textit{E.g.}, the input and output length of the DNN are not mentioned in \cite{bib:DNNwindow}; using a solver, we extrapolated plausible values for these parameters based on the reported computational complexity and hidden layers size. Additional details on this and similar assumptions are reported in \cite{bib:repo}.
\end{itemize}

\subsection{Bit Error Rate}\label{ch:4.2_BER}
\begin{figure}[t]
    \centerline{\includegraphics[width=7cm]{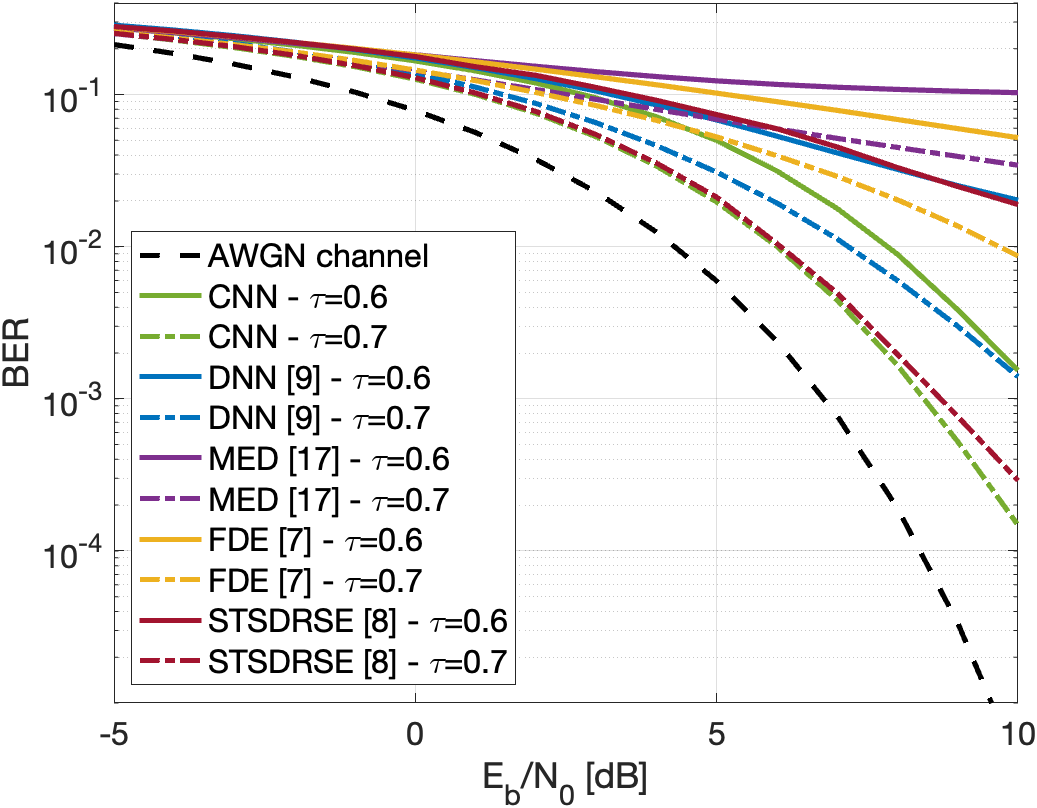}}
    \caption{BER as a function of $E_b/N_0$.}
    \label{fig:ber}
\end{figure}

\noindent The uncoded BER is first evaluated as a function of the $E_b/N_0$ ratio, where $E_b = \frac{E_s}{m}$ is the energy per bit in the FTN signal (Figure \ref{fig:ber}). The black dashed curve reports the theoretical BER with QPSK in AWGN channel (\textit{i.e.}, $\tau=1$). The plot shows that the proposed receiver outperforms all considered benchmarks. At $\tau = 0.7$, the CNN achieves a BER of $10^{-3}$ at an $E_b/N_0$ of 8.5 dB, with a gain of 2 dB against the DNN benchmark and remaining within 2 dB of the ISI-free performance. STSDRSE performs similarly, although taking much longer to converge on the optimization problem solution. Moving to $\tau=0.6$, the stronger ISI widens the performance gap, with the CNN reaching a BER of $1.5\times10^{-3}$ at 10 dB of $E_b/N_0$; on the opposite, the DNN and STSDRSE benchmarks do not reach the $10^{-2}$ threshold. Both compression factors are too strong for FDE, resulting in significantly degraded performance. Without any form of equalization, the ISI-agnostic MED receiver reaches a plateau at $BER = 10^{-1}$ for $\tau=0.6$.
It is worth noting that the DNN benchmark BER curves differ from those reported in \cite{bib:DNNwindow}. We believe the discrepancy is due to different truncation of the channel autocorrelation function samples $\{g_n\}_{n\in\mathbb{Z}}$, leading to different ISI spans $L_I$, as well as our consideration of the boundary conditions in $\mathbf{G}$, which may not have been accounted for in the original work. Nonetheless, our open repository includes the code we used to train the DNN benchmark.

\subsection{Block Error Rate}\label{ch:4.3_BLER}
\begin{figure}[t]
    \centerline{\includegraphics[width=7cm]{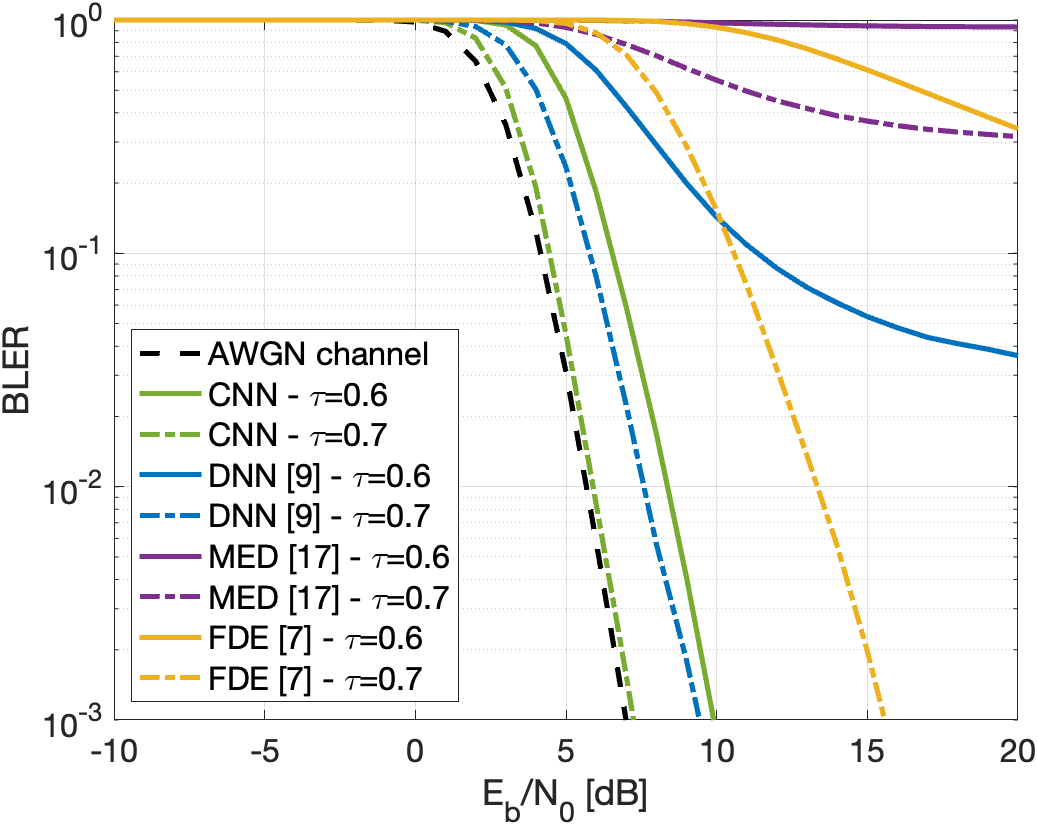}}
    \caption{BLER as a function of $E_b/N_0$ for code rate $R_c=3/4$.}
    \label{fig:bler0.75}
\end{figure}

\begin{figure}[t]
    \centerline{\includegraphics[width=7cm]{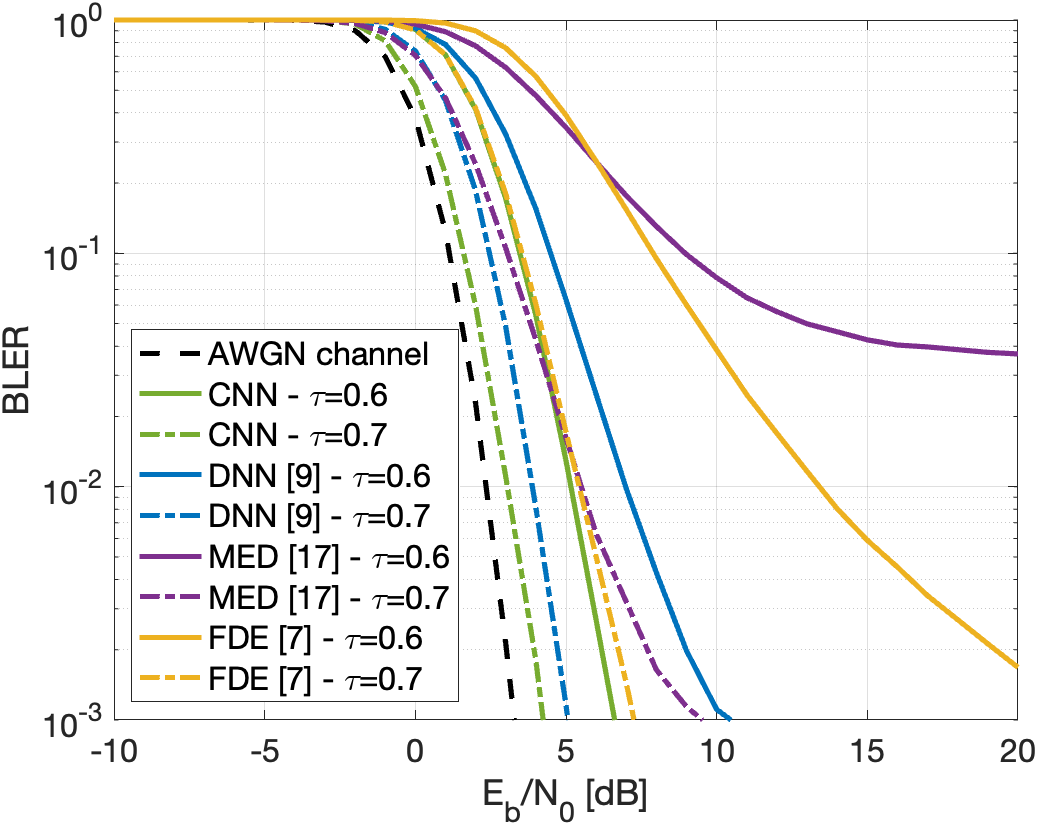}}
    \caption{BLER as a function of $E_b/N_0$ for code rate $R_c=1/2$.}
    \label{fig:bler0.5}
\end{figure}

\noindent We focus now on the BLER, reported in Figures \ref{fig:bler0.75} and \ref{fig:bler0.5} for code rates $3/4$ and $1/2$, respectively. We obtain a theoretical BLER curve by randomly introducing independent bit errors following the theoretical BER in AWGN. At the mildest compression factor, the LDPC decoder is able to correct most of the errors that the CNN was not able to mitigate, leading to a BLER close to ISI-free error rates; the performance degradation at $\tau=0.6$ remains limited, with $BER=10^{-3}$ being achieved at 6.5 dB of $E_b/N_0$ (3dB from AWGN). On the opposite, the DNN requires a 4dB larger $E_b/N_0$ to reach the same threshold, \textit{i.e.}, it achieves a 0.1\% BLER at more than 10dB of $E_b/N_0$. When the code rate is increased to $3/4$, the decoder is limited by the DNN's equalization capabilities, and an error floor is revealed over $BER=10^{-2}$. As in the BER plots, FDE and MED are the least performing receivers, with MED failing to achieve a BLER below 10\%. As previously mentioned, STSDRSE plots are omitted due to the excessive simulation time required to achieve accurate results at high $E_b/N_0$; nonetheless, we note that the algorithm tended to be outperformed by the DNN benchmark for both values of $\tau$.

\subsection{Throughput}\label{ch:4.3_throughput}
\begin{figure}[t]
    \centerline{\includegraphics[width=7cm]{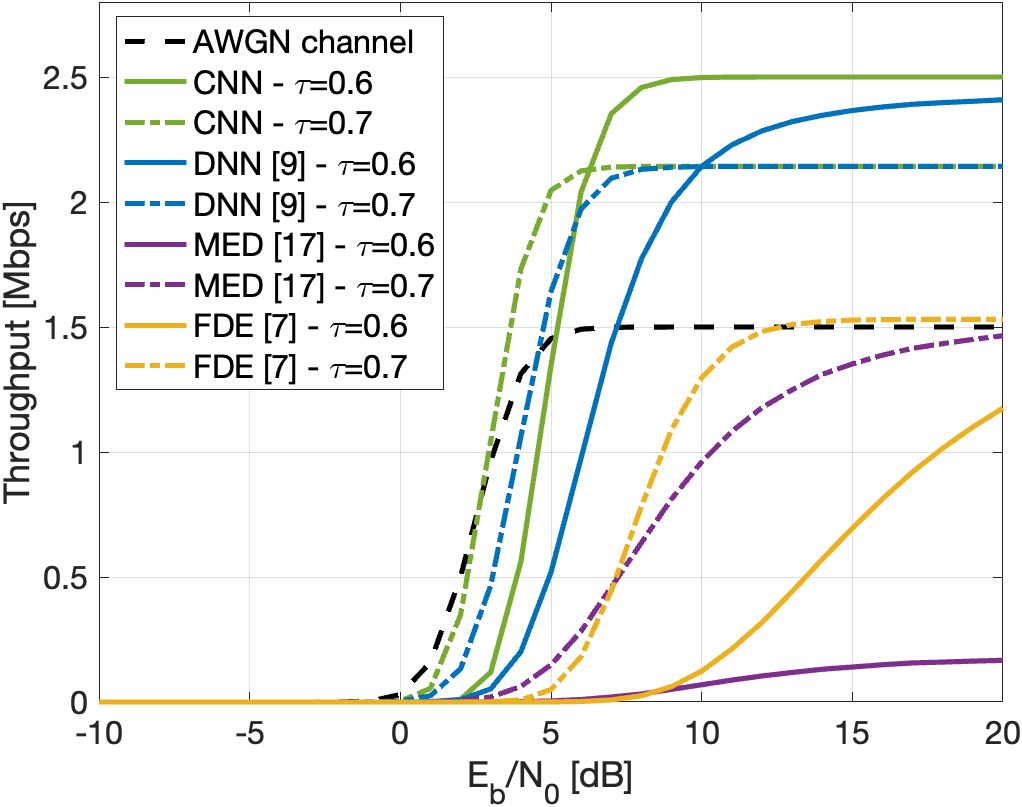}}
    \caption{Throughput as a function of $E_b/N_0$ for code rate $R_c=3/4$.}
    \label{fig:tp0.75}
\end{figure}

\begin{figure}[t]
    \centerline{\includegraphics[width=7cm]{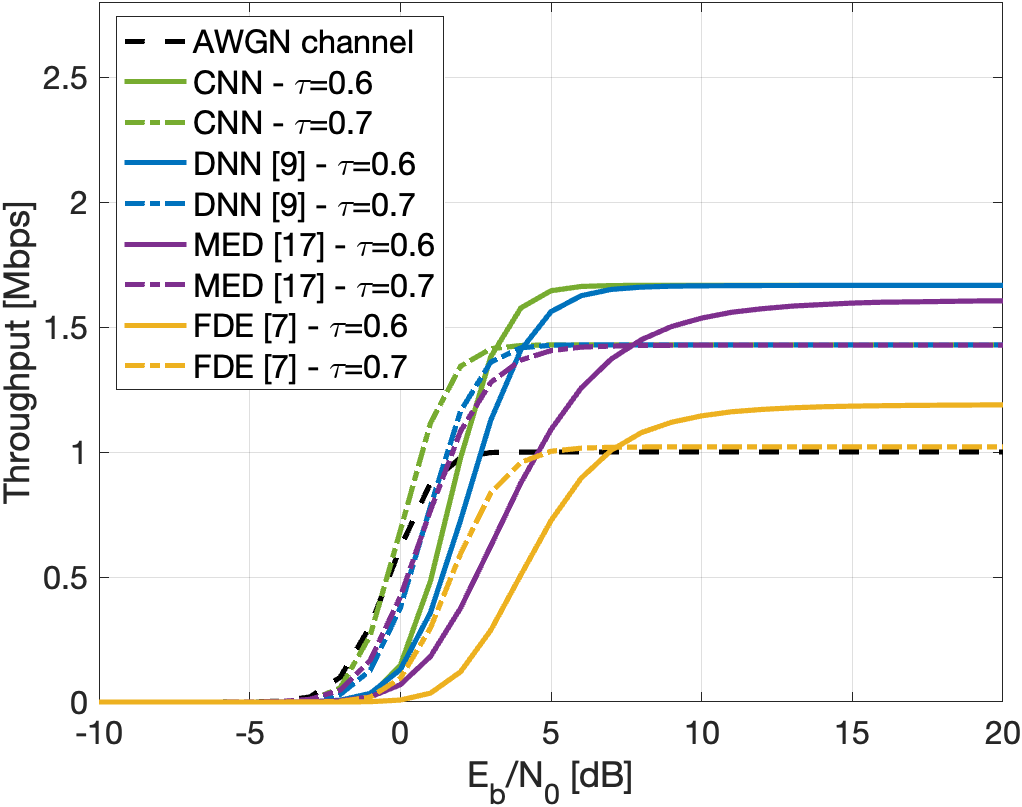}}
    \caption{Throughput as a function of $E_b/N_0$ for code rate $R_c=1/2$.}
    \label{fig:tp0.5}
\end{figure}

\noindent The appeal of FTN lies in its potential TP gains; however, this metric depends not only on the increased symbol rate at the transmitter, but also on the capabilities of the receiver:
\begin{equation}\label{eqn:throughput}
    TP = \frac{m\cdot R_c}{\tau \cdot T_N}\cdot \gamma \cdot (1-BLER),
\end{equation}
where $\gamma = \frac{N_s}{N_s + 2\nu_{FDE}}$ for FDE, accounting for the CP overhead, and $\gamma=1$ for all other receivers. Clearly, $\tau=1$ in the AWGN channel. We investigate the TP gains achieved by the CNN with respect to transmissions at Nyquist rate, comparing such results with the considered FTN benchmarks. Regardless of the code rate, Figures \ref{fig:tp0.75} and \ref{fig:tp0.5} reveal that FTN does not improve the TP at low $E_b/N_0$; this is due to the increased error rate in blocks transmitted at FTN rate. However, the CNN-based receiver surpasses the AWGN TP baseline at lower $E_b/N_0$ values than any other FTN receiver benchmark (2.5dB and -0.5dB for code rate $3/4$ and $1/2$, respectively). At $R_c=1/2$, the proposed receiver reaches its peak TP of 1.47 Mbps and 2.14 Mbps ($\tau=0.6$ and $0.7$, respectively) with a gain of 1-2 dB of $E_b/N_0$ with respect to the DNN benchmark; however, the DNN fails at reaching its peak throughput with high code rate and strong FTN compression due to the BLER floor, while the CNN achieves 2.5 Mbps at an $E_b/N_0$ of 10dB. Notably, the optimal TP curve is achieved by the proposed receiver by initially setting $\tau=0.7$ and $R_c=1/2$, then increasing the coding rate to $3/4$ at 3.5dB of $E_b/N_0$, and eventually switching to $\tau=0.6$ at 6dB of $E_b/N_0$. Indeed, the compression factor $\tau$ can be added as a third dimension to adaptive code and modulation, leading to better TP at any $E_b/N_0$ working point if a powerful FTN receiver, such as the proposed CNN, is implemented.

\section{Computational complexity}
\noindent We showed that CNNs can provide substantial throughput gains in FTN. However, computational complexity is also a key parameter to take into consideration. We evaluate the number of multiply-and-accumulate (MAC) units in a Conv1D layer as $M_{Conv1D} = LDW^{(f)}N^{(f)}$; the additional complexity of BN and LReLU is negligible. Considering the model reported in Figure \ref{fig:CNNscheme}, the total amount of MAC operations amounts to 2.67M, or 53.5k MACs per received symbol. In contrast, the DNN benchmark requires only 8196 multiplications per symbol \cite{bib:DNNwindow}, limiting the number of required MAC units per symbol to such value. Despite the higher computational cost, three key considerations should be taken into account. I) Due to the vast amount of tunable hyperparameters in CNNs, it is possible that lower-complexity configurations yielding similar link-level performance were not fully explored. II) Model compression techniques, \textit{e.g.}, quantization and pruning, though outside the scope of this work, can significantly reduce MAC counts and inference time with negligible performance losses or, in some cases, even marginal gains \cite{bib:modelCompression}. III) Specialized hardware accelerators enable efficient inference with large DL models with small processing time; even with stringent power and size constraints, \textit{e.g.}, on board of small unmanned aerial vehicles, a CNN with 1.1M MACs can run at 139 inferences/s \cite{bib:hardware}. Hence, the adoption of model compression techniques and hardware accelerators makes CNN-based FTN receivers a feasible solution from the computational complexity standpoint.

\section{Conclusions}
\noindent Motivated by recent advances in DL for pattern recognition, this work proposed the application of CNNs to FTN receivers. We trained a CNN with skip connections and evaluated its performance in terms of BER, BLER, and TP against several benchmarks. In an effort to advocate for reproducibility, we open-sourced our code in a public repository. The proposed model achieves BLER levels at less than 1dB and 3.5dB of $E_b/N_0$ from AWGN channel performance for $\tau=0.7$ and $\tau=0.6$, respectively, outperforming all reported benchmarks. These gains directly translate to TP improvements, enabling the full exploitation of FTN's spectral efficiency advantages. Indeed, with $\tau=0.6$ the TP reaches 2.5 Mbps at $E_b/N_0=10$dB, while benchmarks are limited by the strong ISI. Fifty years after Mazo's original paper, thanks to CNNs, FTN has more potential than ever. As current cellular systems employ multi-carrier transmissions, FTN transceivers have recently been proposed for orthogonal frequency division multiplexing (OFDM), \textit{e.g.}, \cite{bib:FTNOFDM}. With the introduction of inter-carrier interference in addition to ISI, we aim at extending the CNN presented in this work to a 2-dimensional case for FTN OFDM receivers and to realistic channel models. Future works will also explore different DL architectures for FTN signaling equalization, \textit{e.g.}, RNNs and Transformers, and techniques to make DL models independent of the modulation order, e.g., modular approaches or output LLR masking.

\section{Acknowledgments}\label{Acknowledgment}
\noindent This work was partially supported by the European Union under the Italian National Recovery and Resilience Plan (NRRP) of NextGenerationEU, partnership of "Telecommunications of the Future" (PE00000001 - program "RESTART"), and by the 6G-NTN project, which received funding from the Smart Networks and Services Joint Undertaking (SNS JU) under the European Union’s Horizon Europe research and innovation programme under Grant Agreement No 101096479. The views expressed are those of the authors and do not necessarily represent the project. The Commission is not liable for any use that may be made of any of the information contained therein.

\bibliographystyle{IEEEtran}
\bibliography{IEEEabrv,IEEEbib}

\end{document}